\documentstyle[twoside,fleqn,espcrc2,epsfig]{article}

\newcommand{\AmS}{{\protect\the\textfont2
  A\kern-.1667em\lower.5ex\hbox{M}\kern-.125emS}}

\title{Heavy sterile neutrinos - what they can be and what they can't}

\author{S.H. Hansen\address{NAPL, University of Oxford, 
Keble road, OX1 3RH, Oxford, UK}
  \thanks{Based on work done in collaboration with A.D. Dolgov, G. Raffelt
and D.V. Semikoz, refs. [3,8,11]. The author is a Marie Curie Fellow.}}
       
\begin{document}

\begin{abstract}
We review current astrophysical bounds on MeV sterile neutrinos, and
then we discuss why a sterile keV neutrino is a natural warm dark
matter candidate.
\end{abstract}

\maketitle

\section{Introduction}
The experimental evidence for neutrino masses and mixing is
overwhelming~\cite{ellis}, and if all the present day experiments are
correct, then there exists at least one sterile neutrino besides the
normal 3 active ones. The mass differences are found to be in the
sub-eV range, but one could naturally imagine additional sterile
neutrinos, and since both masses and mixing angles essentially are
free parameters, one naturally asks the question, which ones are
already excluded - or are there indications that heavy sterile
neutrinos exist?

Let us first review the current bounds on MeV sterile neutrinos, and
then later discuss how a keV sterile neutrino naturally could be
produced in the early universe in just the right amount to be a warm
dark matter candidate.

\section{MeV neutrinos}
Measuring MeV sterile neutrinos is very difficult, and
NOMAD~\cite{nomad} exclude mixing angles for $(\nu_s - \nu_\tau)$
mixing, which go from sin$^2\theta=1$ to $10^{-3}$, when the mass goes
from 10 to 200 MeV. The question is naturally if astrophysical bounds
will overlap this excluded region, and hence completely exclude the
possibility of MeV sterile neutrinos? Let us first consider the bounds
from SN1987A, and then later discuss how the early universe can
provide us with bounds.

The duration of SN1987A gives us an upper limit on the amount of
energy which could have escaped through an ``invisible'' channel, such
as a sterile neutrino carrying away energy and hence shortening the
burst.  The sterile neutrinos are produced in the SN by free neutral
current scattering, and their production rate is
\begin{equation}
\Gamma_s = \frac{1}{2} \, \mbox{sin}^2 2 \theta \, \Gamma_{NC}~.
\end{equation}
Comparing the energy carried away with the observational bound thus
leads to the limit: sin$^22\theta < 3 \cdot 10^{-8}$ for $(\nu_s -
\nu_\tau)$ mixing~\cite{Dolgov:2000pj}. For $(\nu_s - \nu_e)$ mixing
the bound is somewhat stronger~\cite{emix,emix2}: sin$^22\theta <
10^{-10}$. If the mixing angle is too big the sterile neutrino will
never leave the SN, and hence the found bound will not apply, this
translates into: sin$^22\theta > 0.1$.  By comparison with the
terrestrial bound one thus finds a small allowed region for masses
$M<40$ MeV and mixing angles sin$^22\theta > 0.1$.

Big bang nucleosynthesis (BBN) also gives rather strong limits on MeV
neutrinos. Such a neutrino will increase the total energy density,
leading to faster expansion, and one will eventually produce too much
helium.  For a fully thermal species (e.g. the tau neutrino) one finds
that masses $M>0.35$ MeV (corresponding to 0.3 extra neutrino species)
are excluded~\cite{massive}.  Such strong bound can be avoided by
letting the neutrino decay~\cite{dhps}, where one even can reach {\em
minus} 2 extra neutrinos for a 4 MeV tau neutrino decaying with a
lifetime about 0.1 sec into a majoron and an electron neutrino. The
bound will naturally be different for mixed neutrinos. With
non-zero mass and mixing angle the neutrino will decay: $\nu_s
\rightarrow \nu_\tau + l + \bar l$, where $l$ is any of the light
leptons. This translates into a lifetime of the sterile neutrino
\begin{equation}
\tau_s \approx  \frac{1 {\rm sec}}{\left(\frac{M}{10 {\rm MeV}} 
\right)^5 \, {\rm sin}^22\theta}~,
\label{taumix}
\end{equation}
and therefore one also has to include the effects of a non-thermal
spectrum of the electron-neutrino, which enters directly in the
neutron-proton reactions. In particular a bump in the high energy part
of the electron-neutrino spectrum will again lead to overproduction of
helium. First one must write down the momentum dependent evolution
equations for the distribution function of the sterile neutrino,
$f_s$,~\cite{Dolgov:2000pj,Dolgov:2000jw}
\begin{equation}
\partial _x f_s = \frac{f_s^{\rm eq} - f_s}{\tau_s} \, \left[ {\rm Decay}
+ {\rm Collision} \right]~,
\end{equation}
where $x=a\cdot1$ MeV is the expansion parameter of the universe, and
Decay and Collision describe all the possible processes. Similarly one
writes down the equations for the active neutrinos, and for the
temperature evolution. Eventually one find the light element
abundance, and upon comparison with observations one can exclude
regions of the mass-mixing parameter space (see fig.~1).
\begin{figure}[htb]
  \centerline{\hbox{\epsfig{figure=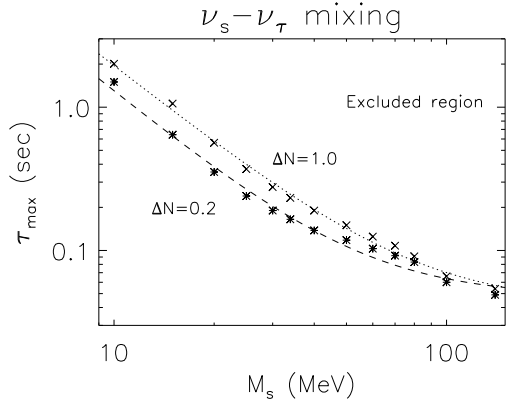,width=0.5\textwidth}}}
  \caption{Maximum lifetime of sterile neutrino mixed with a
  $\nu_\tau$ (or $\nu_\mu$), allowed by BBN, as a function of the
  heavy neutrino mass, both for an optimistic bound, $\Delta N =0.2$,
  and for a conservative bound, $\Delta N = 1.0$ (the figure is taken
  from ref.~[8]).}
\label{fig1} 
\end{figure}

It is now straight forward to translate such bounds into mixing angle
through eq.~(\ref{taumix}), and one finds, that for large mixing
angles this only improves the SN bound slightly, and this only for
small masses, $m=$ few MeV.

\section{Sterile neutrino as warm dark matter}
The sterile neutrinos were absent in the very early universe, and they
were subsequently produced in collisions. The production rate is often
approximated as
\begin{equation}
\frac{\Gamma}{H} = \frac{{\rm sin}^2 2 \theta_M}{2} \left(
\frac{T}{T_W}\right)^3~,
\end{equation}
where H is the Hubble parameter, $\theta_M$ is the mixing angle in
matter, and $T_W$ is the decoupling temperature of the active
neutrinos, $T_W\approx 3$ MeV. In this way one sees, that the
production rate increases as $T^3$ when going to higher
temperatures. The mixing angle is, however, also temperature
dependent~\cite{notraf}, and drops very fast for large temperatures
\begin{equation}
{\rm sin} 2 \theta_M \approx \frac{{\rm sin} 2 \theta}{1 + 0.8 \cdot
10^{-19} T^6 m^{-2}}~,
\end{equation}
where both temperature and mass are measured in MeV. One can plot the
production rate as a function of $1/T$, see fig.~2, and one clearly
sees, that for smaller mixing angles there will be produced fewer
sterile neutrinos, and for smaller masses likewise.
\begin{figure}[htb]
  \centerline{\hbox{\epsfig{figure=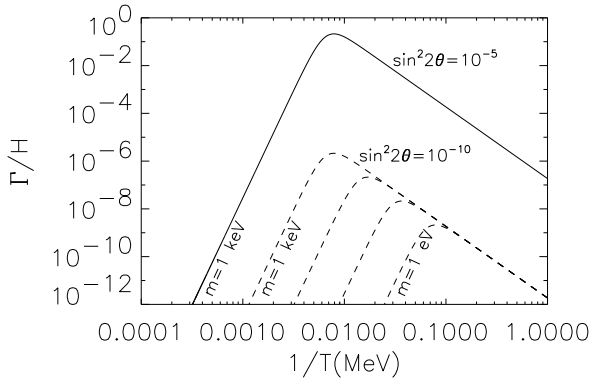,width=0.5\textwidth}}}
  \caption{The production rate $\Gamma/H$ as a function of inverse
  temperature. With a smaller mixing angle there will be produced
  fewer sterile neutrinos, and with a smaller mass likewise.}
\label{fig2} 
\end{figure}
Now one must simply integrate the Boltzmann equation for the
distribution function of the sterile neutrino
\begin{equation}
Hx \partial _x f_s = \frac{{\rm sin}^2 2 \theta_M}{2} \, \Gamma_W \, 
f_\alpha~,
\end{equation}
in order to find the produced sterile neutrinos~\cite{dodwid}. The
simplified approach described above agrees within a factor of 2 with
the more detailed ana\-lysis~\cite{DH}, where also the slight
departure of the sterile neutrino distribution function from the
equilibrium form is described.

Now, taking the Hubble parameter $h=0.65$, and dark matter density
$\Omega_{DM} = 0.3$, one finds the relation~\cite{DH}
\begin{equation}
{\rm sin} ^2 \theta \approx 10^{-14} \cdot \left( \frac{{\rm MeV}}{m}
\right)^2~,
\label{relat}
\end{equation}
which describes a line in the mass-mixing parameter space, where the
sterile neutrino must lie in order to be a good DM candidate.

\subsection{Observational constraints}
We saw above how SN data helped reducing the allowed parameter space
for MeV sterile neutrinos, however, the bound becomes less restrictive
for smaller masses. This is because the mixing angle inside the SN
also is temperature dependent~\cite{raffelt}
\begin{equation}
\mbox{sin}^2 2\theta _M \approx \frac{\mbox{sin}^2 2
\theta}{\mbox{sin}^2 2 \theta + \left( \mbox{cos} 2 \theta + 10^3
\left( \frac{{\rm keV}^2}{m^2} \right) \right)^2}~,
\end{equation}
and hence for masses smaller than $m \sim 40$ keV the limit weakens
substantially.

One must naturally demand that the sterile neutrinos are stable on
time-scales of the universe age, $\tau_s > 4 \cdot 10^{17}$ sec.  From
eq.~(\ref{taumix}) we can thus find the corresponding relationship
between mass and mixing angle: $m($MeV$)^5 \, $sin$^22\theta < 2.5
\cdot 10^{-13}$. A stronger bound is, however, obtained by considering
the radiative decay
\begin{equation}
\nu_s \rightarrow \nu_\tau + \gamma~.
\end{equation}
By comparing with the observations of the diffuse gamma background one
finds the much stronger bound~\cite{DH}
\begin{equation}
m({\rm MeV})^5 \, {\rm sin}^22\theta  < 2.5 \cdot 10^{-18}~,
\end{equation}
which upon comparison with eq.~(\ref{relat}) gives us
\begin{equation}
m < 40 {\rm keV} \, \, \, \, \, {\rm and}  \, \, \, \, \, 
{\rm sin}^2 2 \theta > 10^{-11}~,
\label{bound}
\end{equation}
when the sterile neutrino is mixed with $\nu_\tau$ (or $\nu_\mu$), and
$m < 30$ keV when mixed with $\nu_e$.

\subsection{How to detect or reject?}
The best detection would be searching for a peak in the diffuse gamma
background, which would have the peak energy near $m/2$. A more
careful analysis of the present day data could also strengthen the
bounds found in (\ref{bound}).

The analysis of a future nearby SN will also strengthen the bounds
from the energy loss argument indicated above, and thus potentially
cut away the low mass region. Several other SN aspects will also be
affected by a keV neutrino~\cite{fuller}.

Finally, the real mass of the dark matter particle will be determined
from the analysis of large scale structure formation. At present the
N-body simulations~\cite{nbody,nbody2} point towards a mass about
$0.5$ keV, which in this sterile neutrino picture corresponds to a
mixing angle about sin$^2 2 \theta \approx 10^{-7}$.

\end{document}